**Early Pregnancy Treatment Decisions: Designing Perinatal Pharmacoepidemiology Studies using Real-World Data**

Wood ME,[1] Platt RW,[2,3] Hutcheon JA,[4] Cohen JM,[5] Latour CD,[6] Margulis AV,[7] Petito LC,[8] Grandi SM[9,10]

1 Department of Epidemiology, Gillings School of Global Public Health, University of North Carolina, Chapel Hill, NC, USA
2 Centre for Clinical Epidemiology, Lady Davis Institute, Jewish General Hospital, Montreal, QC, Canada
3 Department of Epidemiology, Biostatistics and Occupational Health, McGill University, Montreal, QC, Canada
4 Department of Obstetrics and Gynaecology, University of British Columbia, Vancouver, Canada.
4 Department of Pediatrics, McGill University, Montreal, QC, Canada
5 Department of Chronic Diseases and Centre for Fertility and Health, Norwegian Institute of Public Health, Oslo, Norway
6 Department of Biostatistics, Epidemiology, and Informatics, University of Pennsylvania, Philadelphia, PA, USA
7 RTI Health Solutions, Barcelona, Spain
8 Division of Biostatistics and Informatics, Department of Preventive Medicine, Northwestern University Feinberg School of Medicine, Chicago, IL, USA
9 Division of Epidemiology, Dalla Lana School of Public Health, University of Toronto, Toronto, Ontario, Canada
10 Child Health Evaluative Sciences, The Hospital for Sick Children, Toronto, Ontario, Canada

**Address for Correspondence:**
Sonia Grandi, PhD
The Hospital for Sick Children
Toronto, ON M5G 0A2
Email: sonia.grandi@sickkids.ca

**Funding**: This work was supported by an award from the International Society for Pharmacoepidemiology. JAH holds a Canada Research Chair in Perinatal Population Health from the Federal Government of Canada. This work was partly supported by the Research Council of Norway through its Centres of Excellence funding scheme, project number 262700.

**Potential COI:**
MEW is a member of the UNC CPE, which receives funding from Abbvie, Astellas, Boehringer Ingelheim, Sarepta Therapeutics, and UCB to support trainee expenses; she has provided methods consultation to Center members on unrelated work.
CDL has received payment from Target RWE, NoviSci, Regeneron Pharmaceuticals, and Amgen for unrelated work.
RWP has received payment from Analysis Group, Daiichi Sankyo, and Merck KGaA for unrelated work.
JMC reports participation in research projects mandated by regulatory authorities and funded by pharmaceutical companies with all funds paid to her institution (no personal fees) and no relation to the work reported in this paper.

**Box 1. Terminology used for data sources**
In this paper, we use the following definitions:

*Administrative health data* is any data source comprising health data primarily collected for purposes other than research, i.e., to support payment, billing, or record-keeping.

*Routinely collected healthcare data* includes data routinely accumulated in the course of contact with the health care system and can include electronic health records and health insurance claims.

*Health insurance claims data* are composed of information, such as diagnosis or procedure codes, about a given healthcare encounter that were paid by the payer, usually a commercial insurance company or a public payer such as Medicaid.

*Electronic health records* are composed of data entered into a patient's medical chart by a clinical practitioner. These data could include diagnosis or procedure codes, measurements such as weight or blood pressure, the results of lab tests, imaging studies, or clinical notes.

*Vital statistics data* include births and deaths that are legally mandated to be reported to a central register, such as birth certificate data in the US.

*Health registry data* include health events such as births, diagnoses, and procedures that encompass care or life events for a population, rather than a frequently-changing group of members (as is the case with health insurance claims).

## ABSTRACT

Research on the use of medications during pregnancy has two primary goals: to detect signals that medications may be harmful to a pregnant individual or fetus, and to support better treatment of pregnant people who require pharmacotherapy. Target trial emulation has been proposed as an approach to estimate the effects of interventions in real world data, with recent extensions to the pregnancy setting. This approach focuses on aligning eligibility and treatment initiation with start of follow up, which is particularly desirable given methodological challenges specific to pregnancy, such as right and left censoring and truncation, competing events, differences in gestational length, and varying etiologically susceptible periods. While previous work on target trial emulation in pregnancy has focused on initiation versus non-initiation of point treatments such as vaccines or antibiotics, this paper focuses on research questions regarding changes to pregestational treatment regimes, and aims to highlight opportunities and approaches to designing studies that align with relevant time points during early pregnancy at which treatment decisions occur in clinical practice. Using the example of treatment for type 2 diabetes mellitus, we review methods for identifying pregnancy episodes in routinely collected healthcare data, introduce possible time zero candidates, and discuss analytic approaches that minimize potential bias due to selection and immortal person time.

## INTRODUCTION

Pregnancy is increasingly common among women who require medications to manage chronic conditions. Changes to pregestational medication regimes may be needed during pregnancy: some medications may negatively affect the pregnancy and the developing fetus, while conversely, under-treatment of some conditions may also cause harm. Ideally, changes to treatment regimes in preparation for pregnancy would be made before conception,[1] but many pregnancies are unintended[2] or not detected until well after conception.[3] As a result, more women must decide in early pregnancy whether to discontinue, continue or switch their pre-pregnancy medications. Clinical guidelines state that such decisions should be informed by a discussion of harms and benefits between the patient and their healthcare providers. Because pregnant women are often excluded from randomized trials, evidence on harms and benefits usually comes from observational studies in routinely collected data.

Studies using routinely collected healthcare data to study medication use in early pregnancy face numerous challenges. Recent work on target trial emulation in pregnancy has discussed the theoretical potential for selection bias in studies of early-gestation medication exposures. This work highlighted the potentially high proportion of pregnancies that end in early loss, which are less likely to be captured in routinely collected data:[4–6] an estimated 20-30% of conceptions[7] and 11-20% of clinically detected pregnancies[8] end in early loss. However, another important issue has received little attention: a frequent disconnect exists between the gestational timeline, which begins at conception, and the timing of real-world treatment decisions, which often occur only after pregnancy is recognized and prenatal care begins.

Using the example of treatment for pregestational type 2 diabetes mellitus (T2DM), we connect elements of the target trial framework with the content and structure of various data sources for studies of medication decisions in early pregnancy. We show how methods for pregnancy identification in real world data may mask issues of selection, illustrate how research questions can be aligned with healthcare encounters when pregnancy-related treatment decisions are made, and discuss approaches to mitigate time-related biases when comparing pregnancy-related treatment decisions.

### Motivating example: treatment of T2DM in pregnancy

For this example, we are interested in whether women with pregestational T2DM taking metformin before pregnancy should continue on metformin alone, add insulin to the existing metformin treatment, or discontinue metformin and switch to insulin at their first prenatal visit or another suitable healthcare encounter. We do not consider discontinuation of all glucose-lowering medications as a possible strategy, as evidence convincingly shows that women with pregestational T2DM benefit from pharmacologic management, and lack of treatment is harmful to both the pregnant person and the developing fetus.[9] Guidelines from the American Diabetes Association and the American College of Obstetricians and Gynecologists recommend insulin as the first line treatment for T2DM in pregnancy, but note that oral antidiabetics such as metformin can be considered if patients are unwilling to switch or unable to tolerate insulin.[9,10] In addition, for patients taking insulin, adjunct metformin treatment can allow reduction of the dose of insulin required to control glycemia. However, questions remain regarding the long-term safety of metformin with respect to neurodevelopment and cardiometabolic conditions among persons exposed to metformin in utero.

To answer this question, we consider a hypothetical target trial among a population of pregnant people with T2DM taking metformin at the time of entry into prenatal care. This hypothetical target trial would use previous pragmatic trials conducted during pregnancy as a framework to inform study design decisions,[11,12] but would be nested in a healthcare system and screen women for eligibility at the time of a first prenatal visit. For illustration, we consider a single time-point intervention, in which the eligible population would be randomly assigned to one of three possible treatment strategies: (1) continuing on metformin alone, (2) adding insulin to metformin, or (3) discontinuing metformin and switching to insulin. Trial participants would then be followed prospectively, from the time of treatment assignment until the occurrence of a study outcome, competing event, or withdrawal from the study; for outcomes in offspring, follow-up would similarly begin at the time of treatment assignment and continue through infancy and childhood. The Table provides an overview of the target trial components, their definitions in a hypothetical trial, and how those components might be emulated in administrative health data.

In the following manuscript sections, we discuss important considerations for emulating this hypothetical trial using routinely collected health care data. We highlight defining and identifying the target population, articulating time zero, and selecting an appropriate analytic approach to avoid immortal time bias.

**1. Identifying and dating pregnancies in routinely collected healthcare data.**

*Identifying pregnancies*

For studies of early pregnancy treatment decisions, the target study population comprises all pregnancies eligible at the time of the specified treatment decision, regardless of subsequent events, such as loss to follow-up or pregnancy loss. However, there is no gold standard source of data that identifies all pregnancies in a population, including those yet unknown to the pregnant persons. Instead, researchers must ascertain a cohort of pregnancies, using a strategy appropriate to the source data for the planned analysis.

Vital statistics records of births or population-based birth registries are a common data source for studies of early-gestation medication use because they cover a clearly-defined population, include high-quality data on some specific key variables such as gestational age at birth, and can often be linked with prescription and claims data.[13–15] However, birth registration is only mandated for pregnancies ending after a minimum gestational age, often ≥ 20-24 weeks; therefore, additional data sources that contain information on pregnancies ending before this gestational age threshold (spontaneous and induced abortions) are needed to create a more complete cohort of clinically recognized pregnancies.

Administrative health data contain diagnosis and procedure codes, prescription data, and other data elements that can be used to identify pregnancies, including non-delivery outcomes in early gestation such as ectopic pregnancy, spontaneous abortions or miscarriages, and induced abortions or terminations.[16,17] Examples of such data include electronic health records and health insurance claims. In these data sources, pregnancies are usually identified using algorithms that identify records with one or more diagnosis or procedure codes for a delivery, live birth, or early pregnancy loss. [17–30] Dispensation records for medications used for care of ectopic pregnancies or medical termination of pregnancy (e.g., mifepristone), pregnancy-specific testing (e.g., non-invasive prenatal genetic testing), or billing records for antenatal care could also be used to enhance identification of pregnancies ending in early gestation or with unobserved outcomes in the source data. Algorithms that capture pregnancies based on prenatal care encounters or pregnancy-specific testing may include pregnancies with no subsequent outcome documentation.[31,32] In addition, there is wide variation in the choice of codes used to define these outcomes, only a limited number of validation studies have established the accuracy of these algorithms, and estimation of gestational age at the time of pregnancy outcome is more error-prone for events occurring earlier in pregnancy.[20,28–30,33,34]

Capturing pregnancy events in administrative healthcare databases relies on the occurrence of a healthcare interaction in which the pregnancy was recorded. The extent to which spontaneous and induced abortions will be captured in healthcare data likely depends in part on the gestational age of the outcome. For example, pregnancy 1 in Figure 1 had a miscarriage at gestational week 4, with no healthcare encounters during pregnancy; this pregnancy is unlikely to be identifiable in healthcare data. In contrast, pregnancy 4 had healthcare encounters in weeks 11 and 13, followed by an early pregnancy loss in week 13; this early pregnancy loss is much more likely to be identifiable in healthcare data. Early pregnancy losses captured in healthcare data are expected to have a later gestational age distribution than that of all losses, and may have other important differences, such as more serious maternal illness leading to greater surveillance.

*Identifying pregnancy start and end dates*

Having identified pregnancies, researchers must then establish the pregnancy start and end dates to assign a gestational age to each health encounter or medication change. The date of conception and the calendar time during which an individual was pregnant are estimated by back-calculating from the gestational age at end of pregnancy or at pregnancy encounters with prenatal timing codes (e.g., ICD-10-CM Z3A codes). Because conception cannot usually be directly observed, gestational age dating is anchored in research and clinical practice to the first day of the last menstrual period, and conception is assumed to occur 2 weeks later. In rare circumstances, including pregnancies conceived with reproductive technologies, the date of conception is known.[34] However, for the majority of recorded pregnancies in administrative health data, date of conception will need to be estimated by subtracting the best estimate of gestational age at birth from the date of delivery (ideally based on ultrasound, otherwise based on last menstrual period estimation), then adding two weeks.

Estimating gestational age for pregnancies that end before the age of birth registration is more challenging, since gestational age is usually less reliable or unavailable for early losses and there may be an unknown interval between the gestational age at fetal arrest of development and the gestational age of the observed miscarriage.[35] Further, current approaches only identify the subset of early pregnancy losses for which health care was obtained, such that less complicated early losses may have less documentation than early losses with complications. In jurisdictions such as the United States, pregnancy terminations might not be covered by some insurance plans and therefore may be missed in claims databases. Including only observed pregnancy outcomes increases confidence that a pregnancy really happened, but risks bias due to conditioning on a future event.[4]

*Choosing a study data source*

The different strategies available for identifying and dating pregnancy cohorts highlight the importance of aligning the choice of study data with the research question. Researchers should carefully consider the availability of key information such as birth census data, early pregnancy loss and termination records, and gestational age dating, as well as data important for their specific research question; for the example of T2DM treatment, lab tests measuring glycemia might be a critical data element. The selection of a data source should be guided by the presence of data elements necessary to answer the research question effectively. If these elements are unavailable and cannot be obtained through data linkages or additional data collection (e.g., questionnaires, chart review), researchers should either select a different data source or revise the research question to ensure it can be appropriately answered with the available data.

**2. Aligning the research question with real-world treatment decisions.**

Studies of medication use in pregnancy have historically focused on exposure, i.e. the risk of the outcome among pregnancies exposed versus unexposed to a medication within a given etiologically-relevant time period, such as the first trimester. However, a study examining the association between first trimester exposure to metformin and risk of congenital anomalies may not provide the most salient information for a pregnant person with T2DM entering prenatal care in the 10$^{th}$ gestational week, who needs to decide whether to change their medication. In this case, the majority of the first trimester has already passed, and while use of an unexposed comparison group may be informative regarding baseline risks, it has limited utility to support the choice of different medications when discontinuation of pharmacotherapy poses risks to both the pregnant person and offspring.

Instead, we propose that researchers should evaluate potential treatment strategies that align with the times at which treatment decisions are being made in clinical care. These strategies can be specified in relation to the existing treatment, and can be broadly categorized as, (1) Continuation: continue with existing treatment; (2) Adding: a new treatment, combined with current treatment, is initiated; (3) Switching: previous treatment is discontinued and a new treatment is initiated; and (4) Discontinuation: the existing treatment is stopped. In our hypothetical trial among pregnant women already taking metformin, the proposed strategies include continuing on metformin alone, adding insulin to the existing metformin treatment, and switching to an insulin-only strategy (Table).

A challenge of using prenatal visits or other healthcare encounters as the start of follow-up is that, unlike in prospectively recruited trials in which treatment assignment is set at baseline, in observational data, patients' subsequent treatment decisions are indistinguishable at baseline. To illustrate this, consider our example of a woman using metformin: the decision to continue metformin alone, add insulin to the existing metformin treatment, or discontinue metformin and switch to insulin would likely occur at the first prenatal encounter, providing a natural anchor for time zero. However, inherent lags in treatment changes based on clinical practice (e.g., clinically-relevant changes in biomarker thresholds), and the ability to observe a treatment change in routinely-collected healthcare data, mean that treatment changes may not occur immediately or be distinguishable at the first prenatal encounter. The availability of longitudinal data for women throughout pregnancy increases the potential for researchers to "look into the future" (i.e., prescription fills after time zero) to define treatment strategies, which may induce immortal time bias.[36,37] Specifying a grace period (e.g., 45 days) within which the treatment strategy should be initiated can minimize this potential bias (see Section 4).

In routinely collected healthcare data, treatment strategies are typically operationalized via prescription order or fill data. This can introduce uncertainty in the precise time at which a treatment began. Assuming that prescriptions are filled for 30-day supplies of a medication, even assuming that prescriptions were taken precisely according to dates of fill and supply, clinical recommendations may include delays in treatment start, gradual titration, adjustment until the desired treatment effect is reached, or changes in response to side effects. Behaviors such as stockpiling, prescription borrowing, missing/forgetting doses, and availability of physician samples mean that medication supply could extend beyond the expected end date. Thus, a clearer articulation of the treatment strategy mimicking the clinical recommendation (e.g., "continue on metformin monotherapy as long as non-fasting blood glucose remains below 120 mg/dL at or below the maximum daily dose, after which insulin should be initiated" versus "continue metformin") allows for clearer operationalization in data, highlights potential limitations of existing data to emulate the desired trial, and is more readily translatable to a clinical audience.

To address these challenges, we recommend: (1) selecting and operationalizing treatment strategies in consultation with clinicians with experience treating the target population, and after conducting detailed descriptive analyses of patterns of prescription fills in the study data, as well as (2) that authors construct illustrations like Figure 2 to provide examples of how observed data would be classified into the treatment strategies under investigation. For example, in Figure 2, classifying Pregnancy 7 as following a "discontinue" versus a "switch" treatment strategy at the first healthcare encounter requires looking into the future, which can cause bias due to immortal person time because a pregnancy must survive longer to be classified as switching versus discontinuing. Clearly articulating treatment strategies that align with clinical practice can aid in defining time periods during which treatment changes would be expected.

### 3. Aligning time zero with relevant time points for treatment decisions

Conception is intuitively appealing as the relevant time zero in pregnancy but, contrasting real-world treatment decisions illustrates that time zero should be anchored on a healthcare encounter. While enrollment in an RCT at conception might be of theoretical interest, limitations of pregnancy tests and delays in pregnancy recognition make it infeasible; these same limitations and delays mean that conception is not the clinically relevant decision point in real patient populations. Prenatal visits, or healthcare encounters during pregnancy, are the interactions with the health care system in which a pregnant person could be randomized, and these visits can serve as natural times at which patients could enroll in the target trial. Further, in our example, they are the natural time points at which decisions to change or continue treatment could occur. Figure 1 illustrates how this approach is necessary to avoid immortal time bias. Consider Pregnancies 2 and 3, which have identical records pre-conception: they both have several interactions with the health system. Both had induced abortions in the 10th gestational week. However, Pregnancy 2 had no encounters between LMP and week 10, while Pregnancy 3 had an encounter at week 6. For Pregnancy 3, the health encounter at week 6 therefore represents a potential treatment decision point and a potential entry into a study, while Pregnancy 2 would not have a natural study entry point. Further, in most of the data sources considered above, no data would be recorded regarding Pregnancy 2 and researchers would not be aware of its existence until the abortion procedure occurred.

Researchers should identify health visits using pregnancy-related codes to ensure that these visits represent visits during pregnancy. The first such visit can be used as the first possible time at which a person might be enrolled.

**4. Approaches to mitigate time-related biases when contrasting treatment decisions.**

Two approaches to address indistinguishable treatment strategies at time zero and minimize immortal time bias include clone-censor-weighting and sequential trials. The clone-censor-weighting approach in non-pregnant populations has been described in detail elsewhere.[38,39] Briefly, this approach avoids immortal time by copying or cloning observations such that there are as many clones as there are possible treatment strategies, and assigning one clone to each strategy. Clones contribute person time to each strategy, beginning at time zero and ending when their assigned strategy is not compatible with the observed data, at which point their follow-up time is censored. Inverse probability of censoring weights are then used to account for the potential selection bias that may occur due to artificial censoring. Figure 3 outlines the use of this approach to examine treatment changes including continuing on metformin alone, adding insulin to the existing metformin treatment, or discontinuing metformin and switching to insulin within a prespecified grace period following a first prenatal encounter. As depicted in the top box, a woman has a first prenatal encounter at 12 weeks' gestation, at which point she is provided with one of three options: within 45 days, she should 1) continue metformin and add insulin; 2) discontinue metformin and initiate insulin; or 3) continue on metformin and do not initiate insulin. Given the inability to distinguish at time zero which of these strategies a woman will eventually follow, this observation is copied three times (or "cloned") and each clone is assigned to one of the three strategies (depicted by the three timelines in the bottom of the figure).

The sequential trial (ST) approach is an alternative approach to mitigate immortal time bias. In a hypothetical ST individuals enter the study when they first become eligible and are randomly assigned to a treatment strategy (e.g., initiate versus do not initiate) at that time. Individuals not assigned to treatment in one trial are then re-randomized at subsequent trials, until they experience the outcome, a competing event, or follow-up ends. Emulation of a ST uses the observed treatment within a short time period (e.g., one week) after each trial's time zero rather than random assignment. The multiple entry points into the trial improve statistical efficiency, while shortening the time intervals being evaluated reduces immortal person time. There are multiple ways to define time intervals for an emulated ST; in pregnancy, gestational weeks have been proposed as reasonable intervals,[37,40] but treatment decision points such as healthcare encounters are also possible.[41]

The ST approach is most useful in settings where the comparison of interest is initiation versus non-initiation of a medication. Figure 4 outlines its use to answer the simplified version of our research question in which the focus is on initiation of insulin versus no initiation, without consideration of metformin. As with the clone-censor-weight approach above, the first time zero is anchored on a prenatal or other health care encounter. At the first trial (termed "Trial 1" regardless of actual gestational week), individuals who fill a prescription for insulin during that gestational week are considered initiators, those without a prescription fill are considered non-initiators, and both groups are followed until the occurrence of a study outcome, a competing event, or the end of follow-up. Initiators in Trial 1 contribute time only to that trial, while non-initiators are eligible for inclusion in Trial 2 and subsequent trials. This approach reduces bias due to immortal person time overall, but residual immortal person time bias may still be present within each trial, since initiators can fill a prescription at any time within the trial interval while non-initiators must complete the full interval without filling a prescription.[35] An alternative approach could instead use prescription fills in the previous time interval to define treatment group, such that medications used in Trial *t* would define treatment group in Trial *t+1*. This approach would reduce immortal person time bias within the index interval, but care is needed to ensure alignment with other eligibility criteria, such as entry into prenatal care or other relevant health care encounters.

There are important considerations for researchers when implementing these approaches. Like all observational studies, these analytic methods do not address confounding bias, highlighting the need for additional methods. Additionally, the ability of these methods to mitigate time-related biases is contingent on the specification of an appropriate time zero, and alignment with initiation of treatment strategies and eligibility, which may not be trivial in administrative claims. Involving researchers with the expertise to implement these designs is critical for success.

## Discussion

The applied and methodologic literature on target trial emulation has expanded in recent years, including examples addressing the specific challenges of pregnancy medication studies.[37,40,42–44] However, prior explanations of how to apply the target trial emulation framework in the context of perinatal pharmacoepidemiology have focused on study questions that contrasted different point treatments, such as COVID-19 vaccination or antibiotic initiation, that are readily distinguishable at baseline. The current work (1) illustrates clinical scenarios in which the relevant clinical questions require contrasting complex treatment regimes for chronic, pregestational conditions, (2) highlights methodologic considerations arising from the limitations of routinely collected health care data, (3) and offers potential strategies for addressing these challenges. Alignment of research questions with opportunities for intervention during pregnancy should result in studies that are better positioned to answer meaningful clinical questions, as in the example of a hypothetical target trial for treatment decisions following initiation of prenatal care among women with T2DM managed with metformin.

This paper addresses study designs for evaluating treatment changes made during pregnancy. For many medications, clinical guidelines recommend making changes to medications before becoming pregnant.[1] Studies on the effects of preconception treatment decisions are important, but face additional challenges in identifying the population at risk, given that not all pregnancies are planned,[2] and not all individuals planning to become pregnant will conceive, or have a pregnancy that is recorded in the health care system, or ends in live birth.[45] Studies estimating the effects of fertility treatments on pregnancy and infant outcomes have proposed the estimation of total effects, including non-conceptions and other competing events in the denominator, and composite outcomes, including non-conceptions plus competing events in the outcome definitions,[46] but others have argued that such estimands are uninterpretable to clinical audiences,[47,48] limiting their practical utility.

We acknowledge, however, that the field of perinatal pharmacoepidemiology is often concerned with early pregnancy exposures that may increase the risk of malformations in offspring, and that regulatory bodies require post authorization safety studies that evaluate and report on malformations as an outcome of interest. Such studies may choose conception as time zero to evaluate risks of exposure to medication in the earliest weeks of pregnancy; however analyses should be presented alongside sensitivity and quantitative bias analyses[23,49] under a range of assumptions about the potential for selection bias.

Evidence-based decision making in pregnancy ultimately requires studies conducted for a range of potential time zeroes, including preconception, conception, as well as those anchored to the earliest healthcare encounters during pregnancy. It is critical that researchers first clearly articulate their research question and then design their study to align with this question. The target trial framework provides a way to do this.

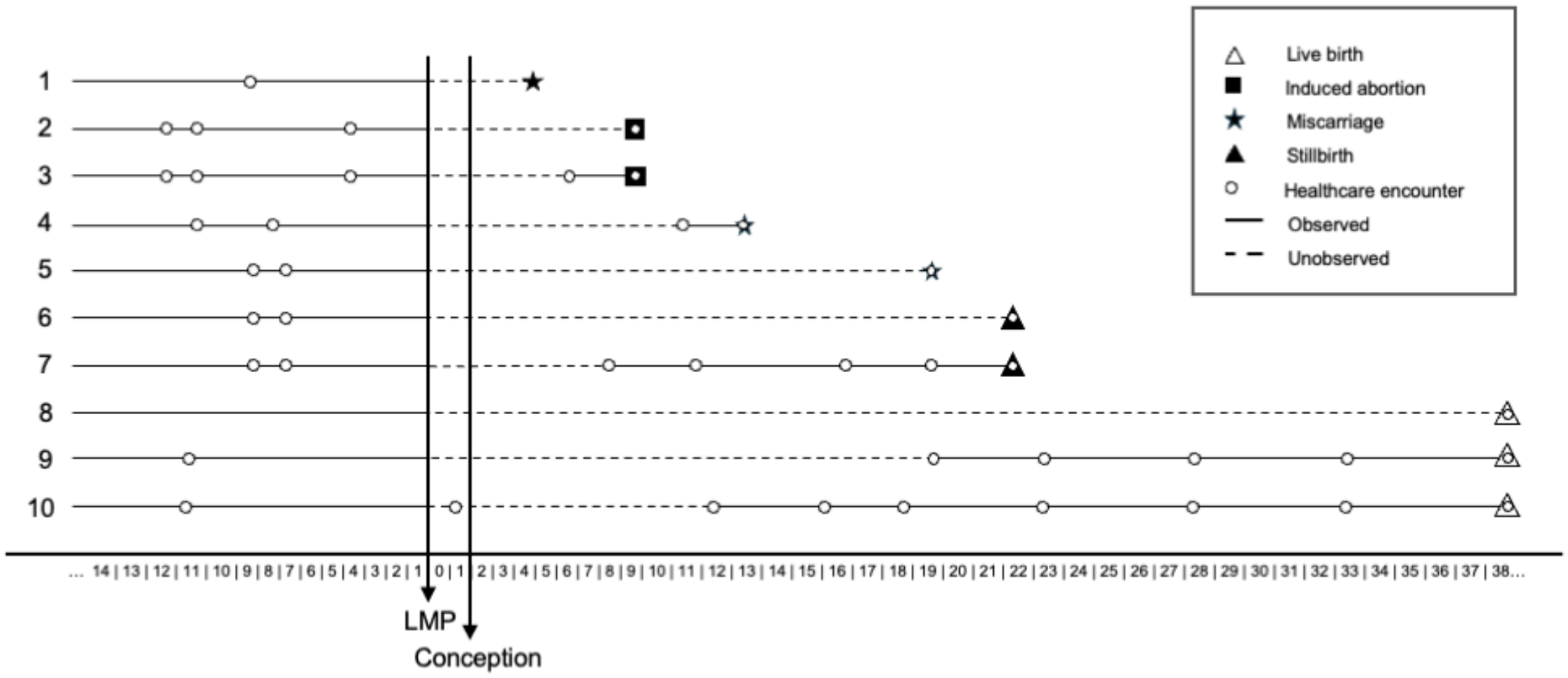

**Figure 1.** 10 pregnancies with a range of outcomes and healthcare encounters, showing all conceptions, with time during which the pregnancy would be observed indicated by a solid line versus unobserved or unknown by a dashed line. Only a subset of pregnancies are expected to appear in administrative health data (e.g., Pregnancy 1 is unlikely to appear in any administrative data source, and whether Pregnancy 2 would be identified likely depends on specific circumstances, such as whether the patient paid for the induced abortion versus having it covered through the payer). Differences in the number and timing of encounters during pregnancy imply varying opportunities for pregnancies to be recorded and treatment decisions to be made.

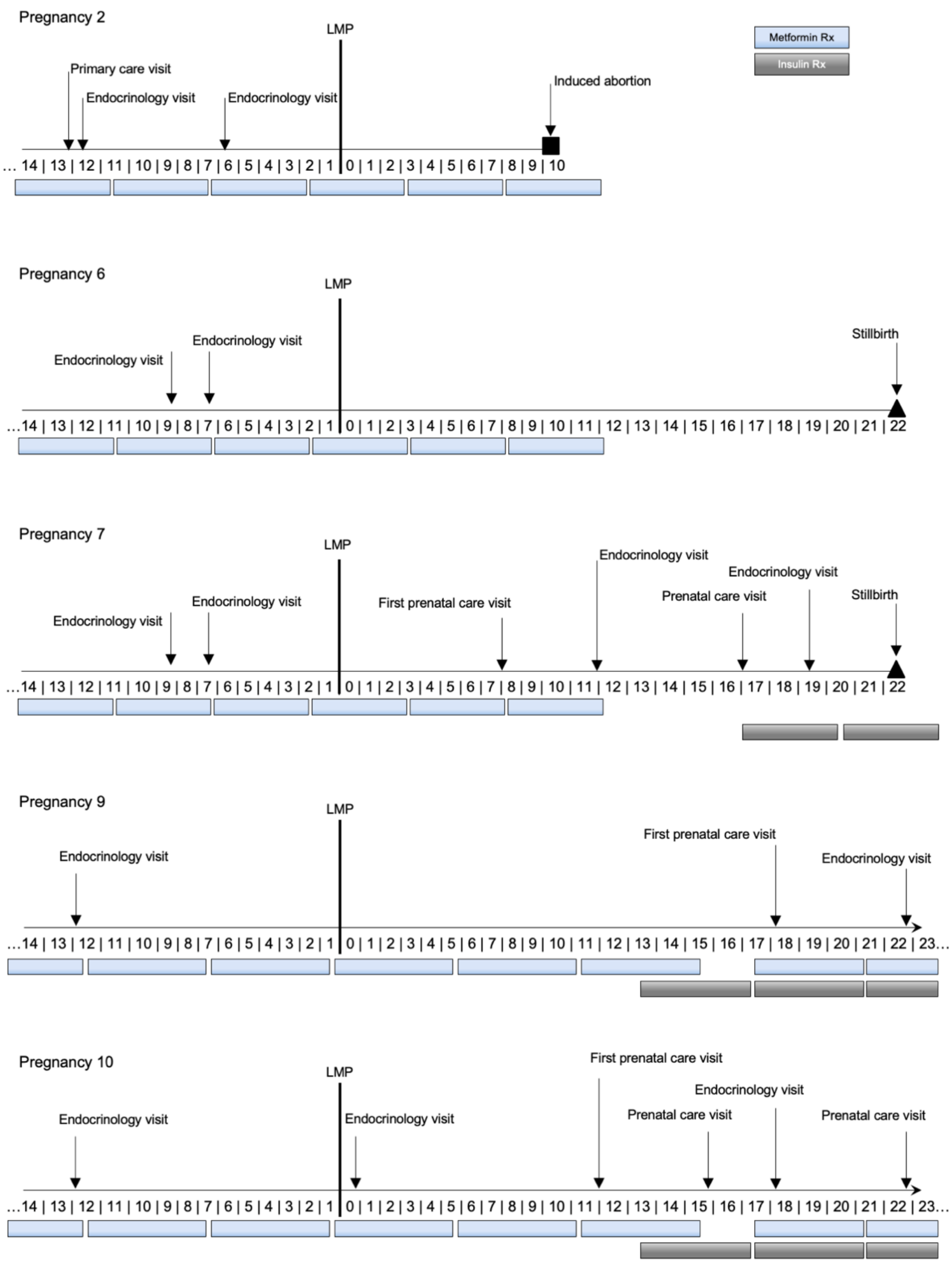

**Figure 2.** Selected pregnancies from Figure 1 are shown in conjunction with prescription fill patterns for metformin (blue boxes) and insulin (gray boxes). The boxes indicate both the date of the prescription fill and the expected coverage based on the supply dispensed.

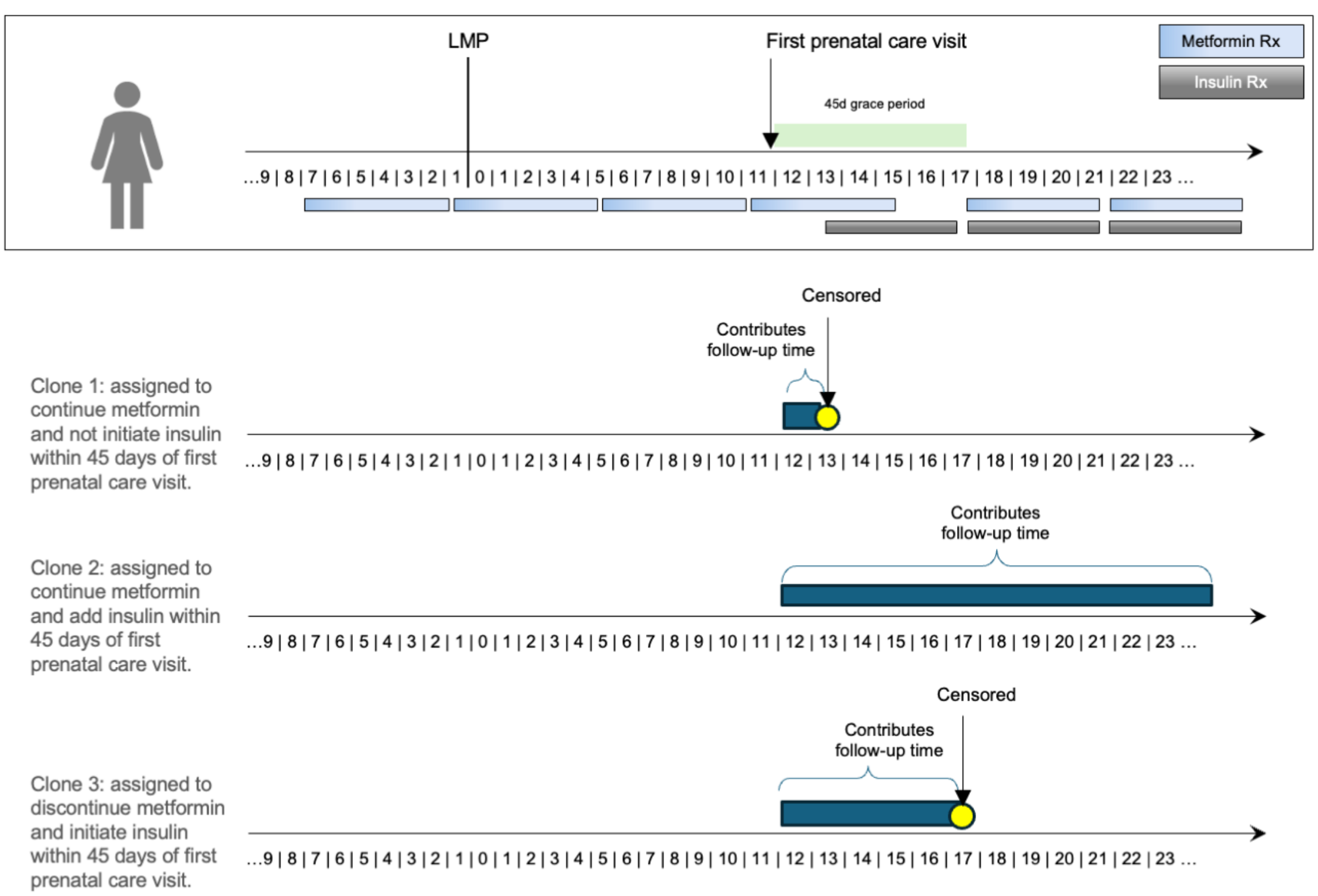

**Figure 3. The clone-censoring-weighting approach to estimate the effect of treatment strategies to manage type 2 diabetes mellitus in pregnancy.**

This figure outlines how to implement the clone-censor-weighting approach in practice using administrative health data. The example depicted in the top box is used to illustrate how this would occur based on the timeline for Pregnancy 10 from Figures 1 and 2. This person presents for their first prenatal visit at week 11, at which point they are eligible for three treatment options: 1) continue on metformin and do not initiate insulin within 45 days, 2) continue metformin and add insulin within 45 days, and 3) discontinue metformin and switch to insulin within 45 days. Using the clone-censoring-weighting approach, this person would be assigned to all three strategies at time zero, in this case, at 11 weeks' gestation. Clone 2's prescription fills are compatible with the observed treatment (continue metformin and add insulin); therefore this clone contributes follow-up time from their prenatal care visit in week 11 until the outcome of interest or the end of pregnancy. Clone 3's prescription fills are consistent with the observed treatment strategy (discontinue metformin and switch to insulin) until week 17 when they fill a prescription for metformin. Therefore, Clone 3 would contribute follow-up time from gestational week 11 to 17, at which time they would be censored. For Clone 1, assigned to continue metformin and not initiate insulin, filling a prescription for insulin in week 13 is a deviation, resulting in their follow-up time being censored at week 13. The final step in this approach involves weighting the person-time for clones not censored in each arm using inverse probability of censoring weights to account for the potential for selection bias induced by the artificial censoring of clones who deviate from their assigned treatment strategies.

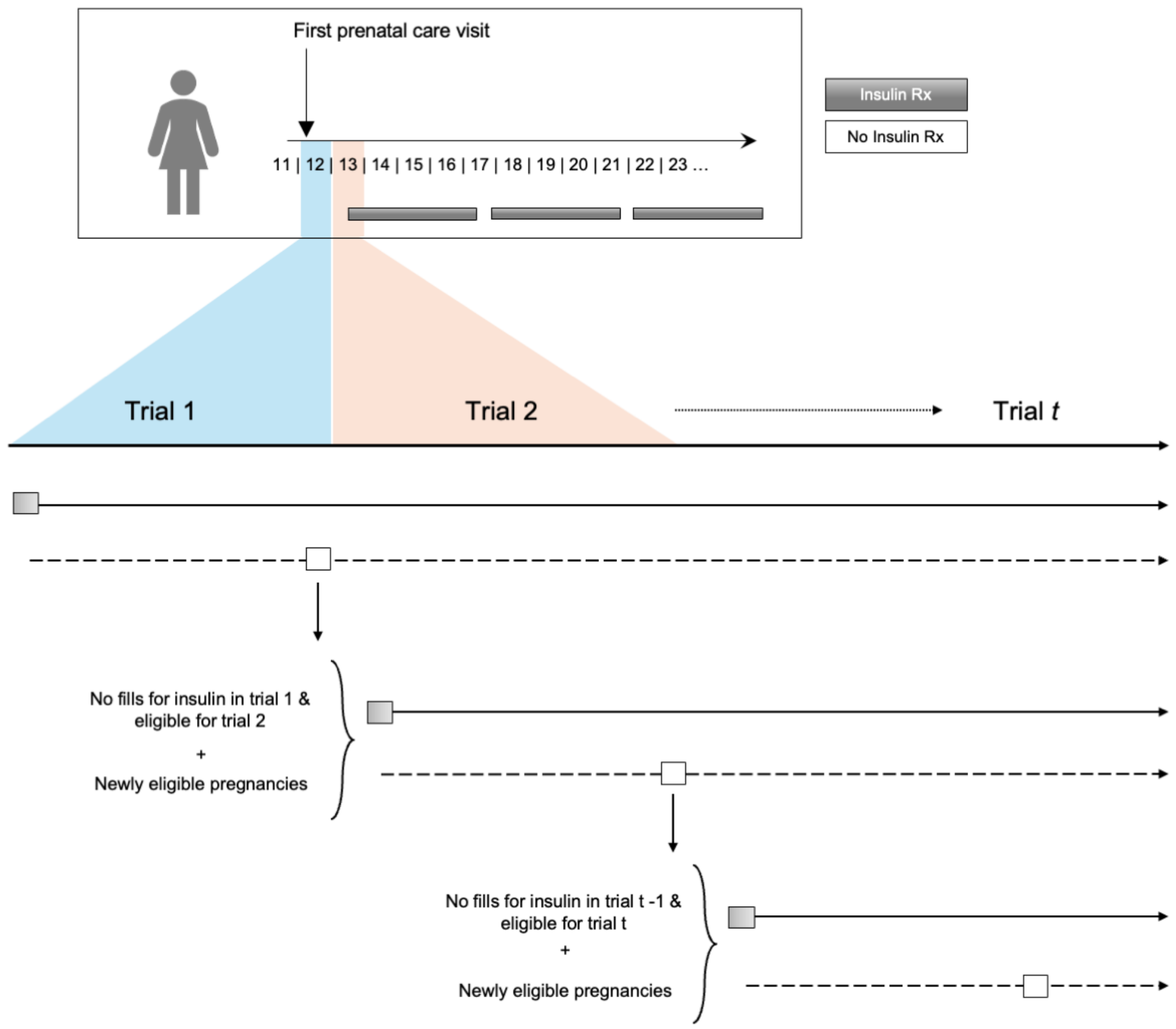

**Figure 4. The sequential trial approach to estimate the effect of initiation of insulin versus no initiation of insulin to manage type 2 diabetes mellitus in pregnancy.**

The top box provides the timeline for Pregnancy 10 from Figures 1 and 2, ignoring metformin prescriptions. This person presents for their first prenatal visit in gestational week 12 and has no prescription fills for insulin during this gestational week; therefore, they are considered an insulin non-initiator in Trial 1 and would be followed until they experience the outcome of interest, experience a competing event, are lost to follow up, or are administratively censored. Non-initiators in Trial 1 are eligible to enter Trial 2 in the subsequent week, together with other non-initiators from week 12 and individuals who are newly eligible, such as those who entered prenatal care in week 13. Because they fill a prescription for insulin in week 13, they are considered an initiator in Trial 2, and are followed until they experience the outcome of interest, a competing event, are lost to follow up, or are administratively censored; they are not eligible for subsequent trials. In a sequential trial of insulin initiation, any individual can contribute to only 1 trial as an insulin initiator but to multiple trials as a non-initiator. An alternative specification of the sequential trial could consider prescriptions in the prior interval as initiation in the index interval: Pregnancy 10 would be considered a non-initiator in Trials 1 and 2, but an initiator in Trial 3 based on their insulin prescription fill in week 13. This latter approach may reduce immortal time within the index interval.

Table. Protocol for a target trial for treatment of T2DM during pregnancy, and its emulation in administrative healthcare data.

| Protocol Component | Description of Target Trial | Description of Emulation in Administrative Healthcare Data |
|---|---|---|
| **Eligibility Criteria** | • Pregnant individuals aged 18-45 years<br>• Ultrasound-confirmed viable singleton gestation<br>• Gestational age < 23 weeks<br>• Intention to continue pregnancy<br>• Diagnosis of T2DM prior to pregnancy<br>• Adherent to maintenance metformin therapy prior to pregnancy with HbA1C <7%<br>• No severe complications of diabetes (e.g., retinopathy, neuropathy), high-risk comorbidities, or contraindications for treatment with insulin. | Same, except:<br>• Conditions will be assessed via structured data (e.g., diagnosis codes in claims data, plus lab results in electronic health records)<br>• Viability and intention to continue pregnancy may not be available |
| **Treatment Strategies** | Participants will follow 1 of 3 possible strategies:<br>1) Continue metformin monotherapy<br>2) Continue metformin and add insulin<br>3) Discontinue metformin and initiate insulin<br><br>Patients will begin following their assigned strategy within 45 days of the index date and continue to follow it throughout pregnancy, with deviations permitted to allow for appropriate clinical management of the pregnancy, or in the case of adverse events. | Same, except:<br>• Treatment will be assessed via structured prescription data (orders and/or fills).<br>• Patient adherence to the strategy will be assessed in the 45-day grace period following the index healthcare encounter. |
| **Assignment Procedures** | Random assignment to exactly one treatment strategy. Patients and providers are aware of treatment assignment. | We will assume that the treatment arms will be exchangeable after adjustment for *a priori* identified baseline confounders. |
| **Follow-up Period** | Patients will be followed from the time of treatment assignment until a study outcome occurs, 6 weeks after the end of pregnancy, or date of loss to follow-up | Same |
| **Outcomes** | Composite of: fetal or neonatal death (fetal loss before 20 weeks, stillbirth ≥20 weeks, neonatal death ≤28 days postpartum), neonatal hypoglycemia, umbilical artery pH <7.05; severe shoulder dystocia, severe hyperbilirubinemia, small- or large-for-gestational age, or birthweight <2500 grams. Outcomes adjudicated by specialists blinded to treatment status. | Same, except conditions will be assessed via structured data (e.g., diagnosis codes). |

| **Causal Contrast** | Intention-to-treat effect; per-protocol effect. | Per-protocol effect |
|---|---|---|

**References**

1. ACOG Committee Opinion No. 762: Prepregnancy Counseling. *Obstetrics & Gynecology* **133**, e78–e89 (2019).
2. Finer, L. B. & Zolna, M. R. Declines in Unintended Pregnancy in the United States, 2008–2011. *New England Journal of Medicine* **374**, 843–852 (2016).
3. Ayoola, A. B., Nettleman, M. D., Stommel, M. & Canady, R. B. Time of pregnancy recognition and prenatal care use: A population-based study in the United States. *Birth* **37**, 37–43 (2010).
4. Latour, C. D. *et al.* Pregnancy identification method as a source of bias in studies of prenatal exposures using real-world data. *Am. J. Epidemiol.* https://doi.org/10.1093/AJE/KWAF260 (2025) doi:10.1093/AJE/KWAF260.
5. Chiodo, S., Tailor, L., Platt, R. W., Wood, M. E. & Grandi, S. M. Emulating a Target Trial in Perinatal Pharmacoepidemiology: Challenges and Methodological Approaches. *Curr. Epidemiol. Rep.* https://doi.org/10.1007/S40471-023-00339-7 (2023) doi:10.1007/S40471-023-00339-7.
6. Suarez, E. A., Landi, S. N., Conover, M. M. & Jonsson Funk, M. Bias from restricting to live births when estimating effects of prescription drug use on pregnancy complications: A simulation. *Pharmacoepidemiol. Drug Saf.* **27**, 307–314 (2018).
7. Wilcox, A. J. *et al.* Incidence of Early Loss of Pregnancy. *New England Journal of Medicine* **319**, 189–194 (1988).
8. Ammon Avalos, L., Galindo, C. & Li, D. K. A systematic review to calculate background miscarriage rates using life table analysis. *Birth Defects Res. A Clin. Mol. Teratol.* **94**, 417–423 (2012).
9. American College of Obstetricians and Gynecologists. Pregestational diabetes mellitus. ACOG Practice Bulletin No 201. *Obstetrics & Gynecology* **132**, e228-247 (2018).
10. American Diabetes Association. 14. Management of Diabetes in Pregnancy: Standards of Medical Care in Diabetes—2021. *Diabetes Care* **44**, S200–S210 (2021).
11. Boggess, K. A. *et al.* Metformin Plus Insulin for Preexisting Diabetes or Gestational Diabetes in Early Pregnancy The MOMPOD Randomized Clinical Trial. *JAMA* **330**, 2182–2190 (2023).
12. Feig, D. S. *et al.* Metformin in women with type 2 diabetes in pregnancy (MiTy): a multicentre, international, randomised, placebo-controlled trial. *Lancet Diabetes Endocrinol.* **8**, 834–844 (2020).
13. Bosworth, O. M. *et al.* Prescription Opioid Exposure During Pregnancy and Risk of Spontaneous Preterm Delivery. *JAMA Netw. Open* **7**, E2355990 (2024).
14. Palmsten, K. *et al.* Oral Corticosteroids and Risk of Preterm Birth in the California Medicaid Program. *Journal of Allergy and Clinical Immunology: In Practice* https://doi.org/10.1016/j.jaip.2020.07.047 (2020) doi:10.1016/j.jaip.2020.07.047.

15. Andrade, S. E. *et al.* Administrative Claims Data Versus Augmented Pregnancy Data for the Study of Pharmaceutical Treatments in Pregnancy. *Curr. Epidemiol. Rep.* **4**, 106–116 (2017).

16. Schummers, L. *et al.* A more accurate approach to define abortion cohorts using linked administrative data: an application to Ontario, Canada. *Int. J. Popul. Data Sci.* **7**, 1700 (2022).

17. Magnus, M. C., Morken, N. H., Wensaas, K. A., Wilcox, A. J. & Håberg, S. E. Risk of miscarriage in women with chronic diseases in Norway: A registry linkage study. *PLoS Med.* **18**, (2021).

18. Ailes, E. C., Simeone, R. M., Dawson, A. L., Petersen, E. E. & Gilboa, S. M. Using insurance claims data to identify and estimate critical periods in pregnancy: An application to antidepressants. *Birth Defects Res. A Clin. Mol. Teratol.* **106**, 927–934 (2016).

19. Ailes, E. C. *et al.* Identification of pregnancies and their outcomes in healthcare claims data, 2008-2019: An algorithm. *PLoS One* **18**, (2023).

20. Andrade, S. E. *et al.* Validation of an ICD-10-based algorithm to identify stillbirth in the Sentinel System. *Pharmacoepidemiol. Drug Saf.* **30**, 1175–1183 (2021).

21. Smith, L. H., Wang, W. & Keefe-Oates, B. Pregnancy episodes in All of Us : harnessing multi-source data for pregnancy-related research . *Journal of the American Medical Informatics Association* https://doi.org/10.1093/jamia/ocae195 (2024) doi:10.1093/jamia/ocae195.

22. Minassian, C. *et al.* Methods to generate and validate a Pregnancy Register in the UK Clinical Practice Research Datalink primary care database. *Pharmacoepidemiol. Drug Saf.* **28**, 923–933 (2019).

23. Huybrechts, K. F. *et al.* Antidepressant use in pregnancy and the risk of cardiac defects. *N. Engl. J. Med.* **370**, 2397–407 (2014).

24. Palmsten, K. *et al.* Harnessing the Medicaid Analytic eXtract (MAX) to Evaluate Medications in Pregnancy: Design Considerations. *PLoS One* **8**, e67405 (2013).

25. Matcho, A. *et al.* Inferring pregnancy episodes and outcomes within a network of observational databases. *PLoS One* **13**, e0192033 (2018).

26. Moll, K. *et al.* Validating Claims-Based Algorithms Determining Pregnancy Outcomes and Gestational Age Using a Linked Claims-Electronic Medical Record Database. *Drug Saf.* https://doi.org/10.1007/s40264-021-01113-8 (2021) doi:10.1007/s40264-021-01113-8.

27. Sarayani, A. *et al.* Impact of the Transition from ICD–9–CM to ICD–10–CM on the Identification of Pregnancy Episodes in US Health Insurance Claims Data. *Clin. Epidemiol.* **Volume 12**, 1129–1138 (2020).

28. Chomistek, A. K. *et al.* Development and Validation of ICD-10-CM-based Algorithms for Date of Last Menstrual Period, Pregnancy Outcomes, and Infant Outcomes. *Drug Saf.* **46**, 209–222 (2023).

29. Lohse, S. R. *et al.* Validation of spontaneous abortion diagnoses in the Danish National Registry of Patients. *Clin. Epidemiol.* **2**, 247–50 (2010).

30. Nordeng, H., Lupattelli, A., Engjom, H. M. & van Gelder, M. M. H. J. Detecting and Dating Early Non-live Pregnancy Outcomes: Generation of a Novel Pregnancy Algorithm From Norwegian Linked Health Registries. *Pharmacoepidemiol. Drug Saf.* **33**, (2024).

31. Giorgio, L. *et al.* Identifying Pregnancies in Population-Based Data Sources: Development and Application of the ConcePTION Pregnancy Algorithm. *Preprint* https://doi.org/10.22541/au.175205384.49458737/v1 (2025) doi:10.22541/au.175205384.49458737/v1.

32. Bertoia, M. L. *et al.* Identification of pregnancies and infants within a US commercial healthcare administrative claims database. *Pharmacoepidemiol. Drug Saf.* **31**, 863–874 (2022).

33. Zhu, Y. *et al.* Validation of claims-based algorithms to identify non-live birth outcomes. *Pharmacoepidemiol. Drug Saf.* **32**, 468–474 (2023).

34. Chiu, Y. H. *et al.* Internal validation of gestational age estimation algorithms in health-care databases using pregnancies conceived through fertility procedures. *Am. J. Epidemiol.* **193**, 1168–1175 (2024).

35. Sundermann, A. C., Mukherjee, S., Wu, P., Velez Edwards, D. R. & Hartmann, K. E. Gestational Age at Arrest of Development: An Alternative Approach for Assigning Time at Risk in Studies of Time-Varying Exposures and Miscarriage. *Am. J. Epidemiol.* **188**, 570–578 (2019).

36. Ukah, U. V. *et al.* Time-related biases in perinatal pharmacoepidemiology: A systematic review of observational studies. *Pharmacoepidemiol. Drug Saf.* **31**, 1228–1241 (2022).

37. Caniglia, E. C. *et al.* Emulating Target Trials to Avoid Immortal Time Bias – An Application to Antibiotic Initiation and Preterm Delivery. *Epidemiology* **34**, 430–438 (2023).

38. Gaber, C. E. *et al.* De-Mystifying the Clone-Censor-Weight Method for Causal Research Using Observational Data: A Primer for Cancer Researchers. *Cancer Med.* **13**, (2024).

39. Maringe, C. *et al.* Reflection on modern methods: Trial emulation in the presence of immortal-time bias. Assessing the benefit of major surgery for elderly lung cancer patients using observational data. *Int. J. Epidemiol.* **49**, 1719–1729 (2020).

40. Hernández-Díaz, S. *et al.* Emulating a Target Trial of Interventions Initiated During Pregnancy with Healthcare Databases: The Example of COVID-19 Vaccination. *Epidemiology* **34**, 1–9 (2023).

41. Wiener, C. *et al.* Comparing causal estimands from sequential nested versus single point target trials: A simulation study. *ArXiv (preprint)* (2026).

42. Huybrechts, K. F., Bateman, B. T. & Hernández-Díaz, S. Modern Evidence Generation on Medication Effectiveness and Safety During Pregnancy: Study Design Considerations.

*Clinical Pharmacology and Therapeutics* vol. 117 895–909 Preprint at https://doi.org/10.1002/cpt.3598 (2025).

43. Schnitzer, M. E., Guerra, S. F., Longo, C., Blais, L. & Platt, R. W. A potential outcomes approach to defining and estimating gestational age-specific exposure effects during pregnancy. *Stat. Methods Med. Res.* **31**, 300–314 (2022).

44. Martinuka, O. *et al.* Target trial emulation with multi-state model analysis to assess treatment effectiveness using clinical COVID-19 data. *BMC Med. Res. Methodol.* **23**, (2023).

45. Wood, M. E. & Edwards, J. K. Conceiving of study designs to evaluate preconception treatment decisions. *JAMA Netw. Open* (2026).

46. Chiu, Y. H. *et al.* The effect of prenatal treatments on offspring events in the presence of competing events: An application to a randomized trial of fertility therapies. *Epidemiology* **31**, 636–643 (2020).

47. Snowden, J. M., Reavis, K. M. & Odden, M. C. Conceiving of questions before delivering analyses: Relevant question formulation in reproductive and perinatal epidemiology. *Epidemiology* **31**, 644–648 (2020).

48. Brown, J. P. & Hernández-Díaz, S. Ending Pregnancy Ends Risks. Consistent Questions, Estimands, Estimates, and Interpretation in the Presence of Competing Events. *Paediatric and Perinatal Epidemiology* Preprint at https://doi.org/10.1111/ppe.70076 (2025).

49. Schnitzer, M. E. & Blais, L. Methods for the assessment of selection bias in drug safety during pregnancy studies using electronic medical data. *Pharmacol. Res. Perspect.* **6**, (2018).